\documentstyle[twoside,fleqn,espcrc2]{article}

\def\pmb#1{\setbox0=\hbox{#1}%
 \kern-.025em\copy0\kern-\wd0
 \kern.05em\copy0\kern-\wd0
 \kern-.025em\raise.0433em\box0 }
\def \bx{{\pmb {$x$}}}
\def \br{{\pmb {$r$}}}

\newcommand{\AmS}{{\protect\the\textfont2
  A\kern-.1667em\lower.5ex\hbox{M}\kern-.125emS}}

\title{Recent Progress in Regge Calculus}

\author{Ruth M Williams\address{DAMTP, Silver Street, Cambridge 
        CB3 9EW, United Kingdom}\address{Girton College, Cambridge
        CB3 0JG, United Kingdom}
        \thanks{This work was supported in part by the UK Particle
                Physics and Astronomy Research Council}}

\begin{document}

\begin{abstract}
While there has been some advance in the use of Regge calculus as a
tool in numerical relativity, the main progress in Regge calculus
recently has been in quantum gravity.  After a brief discussion of
this progress, attention is focussed on two particular, related
aspects.  Firstly, the possible definitions of diffeomorphisms or
gauge transformations in Regge calculus are examined and examples are
given.  Secondly, an investigation of the signature of the
simplicial supermetric is described.  This is the Lund--Regge metric
on simplicial configuration space and defines the distance between
simplicial three--geometries.  Information on its signature can be
used to extend the rather limited results on the signature of the
supermetric in the continuum case.  This information is obtained by a
combination of analytic and numerical techniques.  For the
three--sphere and the three--torus, the numerical results agree with
the analytic ones and show the existence of degeneracy and signature
change.  Some ``vertical'' directions in simplicial configuration
space, corresponding to simplicial metrics related by gauge
transformations, are found for the three--torus.\\

This article is dedicated to Tullio Regge on the occasion of his
sixty--fifth birthday.
\end{abstract}

\maketitle

\section{INTRODUCTION}

1996 is a year in which we celebrate not only the sixty--fifth
birthday of Tullio Regge, but also the thirty--fifth birthday of one
of his brain--children, Regge calculus \cite{Reg61}, a discretization
of general relativity which has provided an amazing source of fascination
and fun for those who have worked on it, as well as the occasional
moment of frustration!

Some of the recent progress in Regge calculus is described elsewhere
in this volume \cite{Imm,Men,Sav}. Rather than attempting a general 
review of other advances, I shall mention a few areas and then concentrate on
one, that of gauge transformations and the simplicial supermetric
\cite{Ham,Har}.

Although much of the current work in Regge calculus involves attempts
at setting up a quantum theory of gravity, there has also been some
progress on the classical front.  An algorithm for the parallel
evolution of sets of disconnected vertices in a spacelike
hypersurface has been formulated \cite{Bar} and should prove an invaluable
tool in numerical relativity, in studies of evolution of model
universes and calculations of gravitational radiation, providing
predictions for measurements by LIGO.  Other classical work includes
an investigation of the use of area variables in four--dimensional
Regge calculus but this issue is not fully resolved as yet.

A very interesting and promising development in the last five years
has been an understanding of the connection between the work of
Ponzano and Regge \cite{Pon} and the state sum or manifold invariant of
Turaev and Viro \cite{Tur}, which uses 6j--symbols for quantum groups to
provide a regularized version of the Ponzano--Regge invariant.  Just
as the semi--classical limit of the Ponzano--Regge invariant is
related to the Feynman path integral for three--dimensional gravity
with the Regge calculus action, the Turaev--Viro invariant is related
to Chern--Simons gravity \cite{Wit}.  Thus it is possible to write down a
finite theory of quantum gravity, with cosmological constant, in three
dimensions \cite{Bar1}.  Barrett and Crane \cite{Bar2} have shown 
recently that the Ponzano--Regge wave function satisfies the Wheeler--DeWitt
equation.  There have been many attempts at extending these ideas to
four dimensions (see for example \cite{Car}
and references in \cite{Wil}) but there are still a number of
open questions and unresolved issues.

The other main area of activity in quantum Regge calculus is in
numerical simulations in two, three and four dimensions \cite{Ham1,Boc,Hol}
In particular there has been extensive investigation of the r\^ole of the 
measure (see also \cite{Men}) and the inclusion of matter represented by
a scalar field.  There is still some controversy over the relationship
between the Regge calculus simulations and the alternative method
known as dynamical triangulations of random surfaces (see eg \cite{Amb}).

Numerical simulations are best guided and complemented by analytical
results.  One of the main tools here is the weak field expansion about
flat space or some other classical solution.  This has been used
recently to study gauge transformations in simplicial gravity \cite{Ham}
and this work will be described more fully in section 2.

The existence of gauge transformations or diffeomorphisms is crucially
important in another area of current research, the study of the
simplicial supermetric.  This is the metric on the space of simplicial
three geometries and a knowledge of its signature is crucial for
determining spacelike hypersurfaces in superspace, important in some
formulations of quantum gravity.  There are so--called ``vertical
directions'' in simplicial configuration space or superspace, which
correspond to metrics which are related by diffeomorphisms and one
objective of current work is to identify these directions.  Sections
3, 4.1 and 4.2 consist of discussions of the known results on the
signature of the supermetric in the continuum, and analytic and
numerical results in the discrete case.

\section{GAUGE TRANSFORMATIONS OR DIFFEOMORPHISMS?}

The first point that needs to be understood here is that there are
differing points of view about how to define gauge transformations or
diffeomorphisms in discrete gravity.  One definition, which tends to
be adopted by those approaching the subject from classical relativity,
is transformations of the edge--lengths which leave the \underbar
{geometry} invariant.  The other, favoured more by those wishing to
use results from lattice gauge theories, is transformations of the
edge--lengths which leave the \underbar {action} invariant.  Some
consider the difference between the two a matter of semantics for,
after all, how would one check that the discrete geometry is left
invariant other than by considering the change in the action?  It is
not quite as simple as that, and depends to some extent on whether one
thinks of the lattice as being embedded in some continuum manifold or
whether one regards the lattice itself as existing independently of
any background embedding.

If one adopts the ``invariance of the geometry'' definition, then a
sensible implementation is to require that all local curvatures (and
hence deficit angles) be unchanged under the transformation of 
edge--lengths.  This is a very strong requirement and will not be met
in general.  The only exception is flat space where, in general, an
infinite number of choices of edge--length will correspond to the same
flat geometry.  In the neighbourhood of flat space there will be
approximate diffeomorphisms \cite{Mor} where, when the edge--lengths change
by order $\epsilon$, the deficit angles will change by smaller than
order $\epsilon$.

On the other hand, in the ``invariance of the action'' definition, it
is easy to imagine changes in the edge--lengths which could decrease
the deficit angles in one region and compensatingly increase them in
another region, producing no overall change in the action.  This 
invariance could even be local in the sense that changes in the lengths 
of edges meeting at one vertex could be made so that the action 
(restricted to scalar curvature, curvature--squared and volume
contributions, say) could be unchanged locally, ie: when evaluated over
the simplices meeting at that vertex.  Before mentioning recent
results demonstrating precisely this locality property, let us
consider in more detail why local gauge invariance is important in 
lattice gravity \cite{Ham2} .  In ordinary gauge theories, local gauge 
invariance plays a central r\^ole as it gives rise to the Taylor--Slavnov 
identities, which ensure for example that the gauge bosons in the theory 
remain massless to all orders of perturbation theory.  Similar results 
hold for lattice regularized versions of these theories.  A small gauge 
breaking in such a lattice theory would invalidate its usefulness as a
representation of the original quantum theory;  in general small gauge
breakings are amplified by loop corrections which pick up
contributions from all momenta and, in particular, short distance
artifacts in the lattice model.  In continuum quantum gravity,
identities analagous to the Taylor--Slavnov identities can be written
down by exploiting the local invariance properties of the functional
integral and, clearly, it is desirable that lattice analogues exist
here as for other gauge theories.  Since it is the lattice action
which appears in the functional integral, invariance of the action is
the relevant consideration here.

In the weak field expansion of Regge gravity about flat space, gauge
transformations, exact zero modes, were found in four dimensions in an
investigations of the lattice propagator \cite{Roc}.  Similar results have
been found in three \cite{Ham3} and in two \cite{Ham} dimensions.  The
form of these zero modes is precisely the discretized form of the
diffeomorphisms in the continuum theory \cite{Ham,Ham3}.  The weak
field expansion about flat space is 
quite relevant for the full quantum theory, because in the
neighbourhood of the ultraviolet fixed point found in three dimensions
\cite{Ham3}
, the average curvature in zero and locally the curvature is small
on the scale of the lattice spacing.

Recent work on edge--length variations about non--flat background
lattices \cite{Ham} indicates that the local gauge invariance is still
there.  It holds also in the presence of a scalar field.

In the remainder of this paper the main emphasis will be on
transformations of the edge--lengths leaving the geometry invariant.
These will be discussed in the context of the simplicial supermetric.
A much more detailed account of this work can be found in \cite{Har}
(see also references therein).

\section{THE CONTINUUM SUPERMETRIC}

The superspace of three--geometries on a fixed three--manifold is
important in various approaches to quantum gravity.  For example, in
Dirac quantization the states are represented by wave functions on
superspace and, in quantum cosmology, such wave functions define
initial and final conditions.

To define the geometry of superspace one needs the notion of
distance.  This is defined on the larger space of three--metrics by
the DeWitt supermetric \cite{Wit1}, the properties of which are
well--understood.  Now the ``points'' of superspace are classes of
diffeomorphically equivalent three--metrics and the DeWitt
supermetric induces a supermetric on this superspace.  The properties
of this induced supermetric are only partially undertood, as we shall
see.

Consider a fixed three--manifold $M$ and let $\mu (M)$ be the space of
three--metrics on $M$, with typical element $h_{ab} (x)$.  The
distance in $\mu$ between two metrics separated by an infinitesimal
displacement $\delta h_{ab}$ is defined by

\begin{equation}
\delta S^2_{\mu} = \int_M d^3 x N(x) \overline G^{abcd} (x) \delta
h_{ab} (x) \delta h_{cd} (x)
\end{equation}

where $N(x)$ is the lapse function and $\overline G^{abcd}$ the
inverse DeWitt supermetric given by 

\begin{eqnarray}
\bar G^{abcd}(x) &=& \frac{1}{2} h^{\frac{1}{2} }(x) [ h^{ac} (x) h^{bd} 
(x) + \nonumber\\
                 & & h^{ad}(x) h^{bc} (x)-2 h^{ab} (x) h^{cd}(x)] 
\end{eqnarray}

This has signature $(-, +, +, +, +, +,)$ and so the signature of the
metric on $\mu$ has an infinite number of both negative and positive
signs.

Denote by Riem$(M)$ the superspace of three--geometries
on $M$.  A supermetric on Riem$(M)$ is induced from the DeWitt
supermetric on $\mu (M)$ by choosing $\delta h_{ab}$ to represent the
displacement between nearby three--geometries.  Since each
three--geometry can be represented by different three--metrics, $\delta
h_{ab}$ is not unique:  for any vector $\xi^a(x)$,

\begin{equation}
\delta h^\prime_{ab}(x) = \delta h_{ab}(x) + D_{(a}\xi_{b)}(x)\ ,
\end{equation}

represents the same displacement in superspace as $\delta h_{ab}$.  Thus
there are so--called ``vertical'' directions in $\mu (M)$, pure gauge
directions, 

\begin{equation}
k^{\rm vertical}_{ab} (x) = D_{(a}\xi_{b)}(x)
\end{equation}

and ``horizontal'' directions, which are orthogonal to all of the
vertical ones, in the metric on $\mu (M)$.

The non--uniqueness of $\delta h_{ab}$ means that there are different
notions of distance between three--geometries in Riem$(M)$.  The
conventional choice is to define $\delta S^2$ to be the minimum value
of $\delta S^2_{\mu}$ for all $\delta h_{ab}$ representing the
displacement between metrics on the nearly three--geometries.  This
means that distance is measured in ``horizontal'' directions in
superspace, which is equivalent to choosing a gauge specified by the
conditions

\begin{equation}
D^b \left(k_{ab} - h_{ab} k^c_c\right) = 0\ .
\end{equation}

In order to define spacelike hypersurfaces in superspace it is
necessary to know the signature of the supermetric.  This is a
non--trivial problem because superspace is infinite--dimensional.  The
known results \cite{Giu,Fri}
include the following:

(i) there is at least one negative direction at each point,
corresponding to constant conformal displacements

\begin{equation}
k_{ab} = \delta\Omega^2  h_{ab}(x)\ .
\end{equation}

This is horizontal because it satisfies the gauge choice above;

(ii)  if $M$ is the three--sphere, then in the neighbourhood of the
round metric, the signature has just one negative direction, the
conformal one, and all other directions are positive;

(iii)  every manifold admits geometries with negative Ricci curvatures;
in the open region of superspace defined by such geometries the
signature has infinite numbers of both positive and negative signs.
This means that, for the sphere, there must be points in superspace
where the metric is degenerate.

These results cover only a small part of superspace, so the procedure
is now to try to obtain more complete information on the signature by
looking at simplicial approximations to superspace based on Regge
calculus.

\section{THE SIMPLICIAL SUPERMETRIC}
\subsection{Analytic Results on the Signature}

A simplicial three--manifold $M$ is constructed from tetrahedra joined
together so that the neighbourhood of each point is homeomorphic to a
region of $R^3$.  A simplicial geometry is ``fixed'' by specifying a
metric with signature (+++);  this is done by assigning (a) values to
the $n_1$ squared edge--lengths;  these must be positive and
satisfy the triangle and tetrahedral inequalities, and (b) a flat
metric, consistent with these values, to the interior of each
tetrahedron.  The region of $R^{n_1}$, with axes the squared
edge--lengths $t^i$, $i = 1,2,..., n_1$, where the inequalities
mentioned in (a) are satisfied, is called simplicial configuration
space, $K(M)$.  Distinct points in $K(M)$, corresponding to different
assignments of edge--lengths in $M$ will, in general, correspond to
distinct three--geometries.  As discussed in section 2, this is not
always true;  the displacements of vertices of a flat geometry in a
flat embedding space give a new assignment of edge--lengths
corresponding to the \underbar {same} flat geometry.  These are the
simplicial analogues of diffeomorphisms and there are also approximate
simplicial diffeomorphisms where the geometry is almost flat locally
\cite{Mor}.  Thus the continuum limit of $K(M)$ is the space of
three--metrics, not the superspace of three--geometries, which is why
$K(M)$ has been called simplicial configuration space rather than
simplicial superspace.

In order to define ``vertical'' and ``horizontal'' directions in
$K(M)$ it is necessary to define a metric:

\begin{equation}
\delta S^2 = G_{mn} (t^\ell) \delta  t^m \delta  t^n\ ,
\end{equation}

where $\delta S^2$ is the infinitesimal displacement between two
points in $K(M)$.  The supermetric $G_{mn}$ can be induced from the
DeWitt supermetric on the space of continuum three--metrics as follows.
A simplicial geometry in $K(M)$ can be represented (not uniquely) by a
point in $\mu (M)$ corresponding to a piecework flat metric, and a
displacement $\delta t^m$ in $K(M)$ can be represented by a
perturbation $\delta h_{ab}$ in $\mu (M)$.  Then we define

\begin{eqnarray}
G_{mn} (t^\ell) \delta t^m \delta t^n &=& \int\nolimits_M d^3 x\, N(x) 
\,\bar G^{abcd}(x) \nonumber\\
& & \times\delta h_{ab}(x) \delta h_{cd}(x)
\end{eqnarray}

where $\overline G^{abcd}$ is evaluated at the the piecewise flat
metric. To make this definition meaningful we need to specify the
following:

(i)  the value of $N(x)$.  Take $N (x) = 1$;

(ii)  the gauge inside each tetrahedron.  We call this choice the
``Regge gauge freedom'' and a natural choice is to take $\delta
h_{ab}$ to be constant inside each tetrahedron:

\begin{equation}
D_c \delta h_{ab} (x) = 0\ , \quad {\rm inside\ each}\quad
 \tau\in \Sigma_3\ ,
\end{equation}

where $\sum_3$ is the set of tetrahedra in $K(M)$.

Of course, there may still be variations of the edge--lengths which
preserve the geometry, so this type of gauge freedom remains.

Evaluation of the integral above leads to the expression

\begin{eqnarray}
G_{mn}(t^\ell)\delta t^m\delta t^n = &\sum\limits_{\tau\in\Sigma_3}&
V(\tau)\{\delta h_{ab} (\tau) \delta h^{ab}(\tau) \nonumber\\
         & & -\left[\delta h^a_a(\tau)\right]^2\}
\end{eqnarray}

where $V(\tau)$ is the volume of tetrahedron $\tau$.

The relation between $h_{ab}$ and the squared edge--lengths
$t_{ab}$, linking vertices $a$ and $b$, may be obtained by picking
a vertex 0 in a tetrahedron and taking basis vectors along the edges
from that vertex.  Then

\begin{equation}
h_{ab}(\tau) = \frac{1}{2} \left(t_{0a}+t_{0b}-t_{ab}\right)
\end{equation}

and hence

\begin{equation}
\delta h_{ab} (\tau) = \frac{1}{2} \left(\delta t_{0a} + \delta
t_{ab} - \delta t_{0b}\right) \ .
\end{equation}

This fixes the Regge gauge for each tetrahedron.

We now use the expression \cite{Har1}

\begin{equation}
V^2(\tau) = \frac{1}{(3!)^2} {\rm det} [h_{ab}(\tau)]\ .
\end{equation}

and consider perturbations $\delta t^l$ corresponding to $\delta
h_{ab}$.  Expansion of 

\begin{equation}
V^2 (t^l + \delta t^l) = {1 \over (3!)^2} {\rm det} (h_{ab} + \delta
h_{ab})
\end{equation}

via

\begin{equation}
V^2 (t^l + \delta t^l) = {1 \over (3!)^2} {\rm exp} [Tr \log (h_{ab} + 
\delta h_{ab})] 
\end{equation}

leads, at first order, to

\begin{equation}
\delta h^a_a(\tau) = \frac{1}{V^2}\ \frac{\partial V^2(\tau)}{\partial
t^m}\ \delta t^m\ ,
\end{equation}

and, at second order, to

\begin{equation}
G_{mn} (t^\ell) = -\sum\limits_{\tau\in\Sigma_3} \frac{1}{V(\tau)} 
\ \frac{\partial V^2(\tau)}{\partial t^m\partial t^n}\ .
\end{equation}

This is the expression written down by Lund and Regge \cite{Lun}
for the
metric on $K(M)$.  It can be shown that this gives the shortest
distance between simplicial three--metric among all choices of Regge
gauge which vanish on the triangles.  It is not exactly "horizontal"
in the sense of the continuum because of the possibility of simplicial
diffeomorphisms.

We are particularly interested in the signature of the Lund--Regge
supermetric $G_{mn}$ and we give here several analytic results which
give limited information about it.

(i)  The conformal direction is always timelike.  Suppose that

\begin{equation}
\delta t^m = \delta{\Omega^2} \ t^m \ .
\end{equation}

Now $V^2$ is a homogeneous polynomial of degree three in $t^l$, so
Euler's theorem applied to the expression for $G_{mn}$ above leads to

\begin{equation}
G_{mn} (t^\ell) t^m t^n = - 6V_{\rm TOT} (t^\ell) < 0\ .
\end{equation}

Also, the conformal direction is orthogonal to any gauge direction
$\delta t^2$ because 

\begin{equation}
G_{mn} = t^m \delta t^n = -4 {\partial V \over \partial t^n} {\rm TOT}
\delta t^n = -4 \delta V_{{\rm TOT}}
\end{equation}

which is zero for a gauge transformation (by which we mean here any
change in edge--lengths which does not change the geometry).

(ii)  There are at least $n_1 - n_3$ spacelike directions where $n_3$
is the number of tetrahedra in $K(M)$.  This follows from showing that

\begin{eqnarray}
G_{mn} \delta t^m \delta t^n &=& -4 \sum_\tau \frac{1} { V( \tau ) } 
\frac{\partial V(\tau)} {\partial t^m} \frac{\partial V ( \tau ) } 
{\partial t^n} \delta t^m \delta t^n \nonumber\\ 
& &+\sum_\tau V(T) \delta h_{ab} (\tau) \delta h^{ab} (\tau) 
\end{eqnarray}

which is non--negative if

\begin{equation}
{\partial V(\tau) \over \partial \tau^m} \delta t^m = 0 \qquad {\rm
for \ all}\ \tau \ .
\end{equation}

Therefore variations of the edge--lengths leaving the volumes of all
tetrahedra unchanged correspond to spacelike directions. There are
$n_3$ conditions on $n_1$ edge--length changes so there are at least
$n_1 - n_3$ independent spacelike directions.

(iii)  Diffeomorphism modes:  consider a flat simplicial geometry
embedded locally in $R^3$, with vertices at $\bx_A, A =
1,2,...,n_0$. Displacements of these vertices through $\delta \bx_A$
produce changes in the edge--lengths $\delta t^m$ which correspond to
gauge directions since the flat geometry is unchanged.  It can be
shown (see \cite{Har} for details) that

\begin{eqnarray}
\delta S^2 &=& 2 {\partial V_{TOT} \over \partial t_{CD}} \delta
\bx^2_{CD}\nonumber\\
& &-\sum_{\tau} {2 \over V(\tau)} \Biggl( {\partial
V(\tau) \over \partial t_{CD}} \bx_{CD} \cdotp \delta \bx_{CD}
\Biggr)^2
\end{eqnarray}

Although the second term gives a negative contribution and the first
does not appear to have a definite sign, I conjecture that gauge modes
are positive (see further comments on this in the numerical section).

\subsection{Numerical Results on the Signature}

The limited analytic information on the signature of the simplicial
supermetric described above may be checked and supplemented by
numerical evaluation of the supermetric over the whole of simplicial
configuration space or at least a non--trivial region of it.  The
general method employed here is as follows:

(i)  choose a simple triangulation of a chosen manifold;

(ii)  assign edge--lengths (consistent with the triangle and
tetrahedral inequalities);

(iii)  calculate the Lund--Regge supermetric $G_{mn}$;

(iv)  find the eigenvalues of the matrix $G_{mn}$:

(v)  count the numbers of positive, negative and zero eigenvalues to
determine the signature;

(vi)  update the edge--lengths and repeat the procedure.

We now describe the results for two manifolds, $S^3$ and $T^3$.

\subsubsection{The three--sphere}

We chose the simplest triangulation of $S^3$, the surface of a
four--simplex, which has 5 vertices, 10 edges, 10 triangles and 5
tetrahedra.  Thus, in this case, simplicial configuration space is
10--dimensional, each point in it corresponding to an assignment of 10
squared edge--lengths.

Since the signature is scale invariant, the longest edge can be fixed
at limit length.  To sample the whole space by dividing the limit
interval into 10, it looks at first as though there will be 10$^{10}$
points to investigate.  However, the triangle and tetrahedral
inequalities reduce this to 102,160 points.  The results on the
signature are the same everywhere; the signature is
$(-,+,+,+,+,+,+,+,+,+)$.  It is easy to see that this is consistent with
the analytic results. 

The results on the eigenvalues of $G_{mn}$ are also consistent with
symmetry considerations. The symmetry group of the triangulation is
$S_5$, the permutation group on five vertices.  For the case of equal
edge--lengths in the triangulation, the eigenvalues of $G_{mn}$ may be
classified by the irreducible representations of $S_5$ and their
degeneracies given by the dimensions of those representations.  This
is because a permutation of the vertices corresponds to a matrix
acting on the 10--dimensional space of edges.  These matrices give a
10--dimensional reducible representation of $S_5$ which can be
decomposed into irreducible representations as 

\begin{equation}
10 = 1 + 4 + 5 \ .
\end{equation}

This can be compared with the numerical results when all the $t_i$ are
one.  The eigenvalues of $G_{mn}$ are $-1/\sqrt 2$ with degeneracy 1,
$1/3 \sqrt 2$ with degeneracy 4, and $5/6 \sqrt 2$ with degeneracy 5.
In general, these degeneracies are broken by departure from the
completely symmetric assignment of edge--lengths.

\subsubsection{The three--torus}

To obtain a triangulation of $T^3$, we take a lattice of cubes, with
$n_x, n_y$ and $n_z$ cubes in the $x-, y-$ and $z-$directions
respectively.  Each cube is divided into 6 tetrahedra by drawing in
face diagonals and a body diagonal \cite{Roc}.  Then there are $n_0 = n_x
n_y n_z$ vertices, $7n_0$ edges, $12n_0$ triangles and $6n_0$
tetrahedra, and simplicial configuration space is $7n_0$-dimensional.
We considered two particular flat triangulations, one with
right--angled tetrahedra and one with isosceles tetrahedra (see \cite{Har}
for details).  The computations ranged from $3 \times 3 \times 3$
vertices and 189 edges to $6 \times 6 \times 7$ vertices and 1764
edges.  Unlike the three--sphere case, the numbers of edges, even in
the smallest case, mean that it is impossible to sample the whole of
simplicial configuration space, so the investigation was done in two
ways, firstly by making random variations in the edge--lengths about
flat space and secondly by making perturbations along the flat--space
eigenvectors $v^j$:

\begin{equation}
t^i_{{\rm new}} = t^i_{{\rm flat}} + \epsilon v^j \ ,
\end{equation}

for all $t^i$ and each $v^j$ in turn, with $\epsilon$ small.

The results described here are for the 189--dimensional case.  For
flat space, the signature has 13 negative signs and 176 positive ones.
This changes away from flat space and there are even some zero
eigenvalues which are presumably finite analogues of the infinite
number of non-gauge directions predicted by Giulini \cite{Giu}
for open regions with negative Ricci curvature.

To understand the degeneracy of the eigenvalues, we need to look at
the symmetry group of the lattice.  For the $3 \times 3 \times 3$
case, this group has a subgroup of translations (mod 3) complemented
by a subgroup $Z_2 \times S_3$ (corresponding to ``parity''
transformations and permutations of the axes).  The 189--dimensional
reducible representation of this group decomposes as

\begin{eqnarray}
189 = & & 3 \times (1_1) + 2 \times (2_1) + 3 \times (2_2) \nonumber \\
&+& 2 \times (4) + 5 \times (6_1 + 6_2 + 6_3+ 6_4) \nonumber\\
&+& 2 \times (6_5 + 6_6 + 6_7 + 6_8) 
\end{eqnarray}

where $2_1$ is the first irreducible representation of dimension 2,
and so on.  The multiplicities of the eigenvalues of $G_{mn}$ for the
right--tetrahedron lattice agree exactly with this complicated
decomposition provided that the two apparent multiplicities of 8 are
interpreted as 6 + 2, and the two multiplicities of 3 are interpreted
as 2 + 1.  We do not understand such accidental degeneracy, but it has
been observed elsewhere, for example for some triangulations of $CP^2$
\cite{Har2,Har3}.

To investigate the existence of gauge modes, the edges are varied in
the directions of the eigenvectors as already described and the new
deficit angles calculated to see whether the geometry remains flat.
For the isosceles tetrahedral lattice, the results are that the
eigenvectors with eigenvalues $\lambda = \textstyle {{1 \over 2}}$
appear to be approximate diffeomorphisms, since the deficit angles are
of order $\epsilon^3$ and, for an $n_0 = n \times n \times n$ lattice,
there are $6n - 4$ eigenvectors corresponding to this eigenvalue.

The conjecture about positive eigenvalues for the gauge modes has not
been proved but has support from a number of special cases for the
right--tetrahedral lattice, where it reduces to

\begin{equation}
\delta S^2 = \sum_{CD} \delta \bx^2_{CD} - {1 \over 12} \sum_{\tau}
\Biggl[ \sum_{{C, \atop \displaystyle{\epsilon}} {D \atop
\displaystyle{\tau} }} \, \bx_{CD} \cdotp
\delta \bx_{CD} \Biggr]^2
\end{equation}

Thus $\delta S^2$ is positive in the following cases:

(i)  if just one vertex moves through $\delta \br$,

\begin{equation}
\delta S^2 = 8 \delta \br^2
\end{equation}

(ii)  if the edges in only one co--ordinate direction change,

\begin{equation}
\delta S^2 \geq {1 \over 2} \sum \delta \bx^2_{CD}
\end{equation}

(iii)  if all the $\delta \bx_{CD}$'s have the same magniture $\delta
L$, 

\begin{equation}
\delta S^n > (\delta L)^2 (n_1 - {3 \over 4} n_3) > 0
\end{equation}

since $n_1 > n_3$ \ . 

Let us briefly summarize the numerical results of this section.  For
triangulations of both manifolds there are negative modes, including
the conformal direction.  For the five--cell triangulation of $S^3$
there is a single negative mode, but preliminary results on the
600--cell triangulation indicate a total of 92 \cite{Bro}.  In all the cases
studied there are at least $n_1-n_3$ positive modes.  The gauge modes
found for $T^3$ are positive but we have no proof that this is
generally true.  The results for $T^3$ show that the supermetric can
become degenerate and the signature changes as simplicial
configuration space is explored.

The main advantage of the discrete approach to the supermetric is that
the continuum infinite dimensional superspace is reduced to simplicial
configuration space which is finite dimensional but preserves elements
of both the physical degrees of freedom and the diffeomorphisms.  For
larger and larger triangulations, more and more aspects of both should
be recovered.  

One obvious line of further study is to understand the vertical and
horizontal directions in simplicial configuration space and relate
them to the vertical and horizontal directions in the continuum.
Another avenue is to extend the work on $S^3$ to a triangulation with
an arbitrarily large number of vertices, which would then encode all
the dynamic and gauge degrees of freedom, unlike the five vertex model
studied here, which is too small to exhibit the expected number of
(approximate) gauge modes (the number of edges 10 is less than $3n_0$
in this case).  Of course, the methods described here could also be
applied to triangulations of other manifolds.

\section{ACKNOWLEDGEMENTS}

I would like to thank the organizers of the Second Meeting on
Constrained Dynamics and Quantum Gravity, Santa Margharita Ligure,
Italy, September 1996 for carrying out their tasks with such wisdom
and efficiency.  I would also like to thank my collaborators Herbert
Hamber, James Hartle and Warner Miller for their major contributions
to the work described here.  


\end{document}